\documentclass[pra,twoside,showpacs,superscriptaddress,twocolumn]{revtex4}

\usepackage{latexsym}
\usepackage{graphics}

\begin{document}

\title{Programmable discriminator of coherent states -- experimental realization}

\author{Lucie Bart{\accent23 u}\v{s}kov\'{a}}
\affiliation{Department of Optics, Palack\'y University,
     17.~listopadu 50, 772\,00 Olomouc, Czech~Republic}

\author{Anton\'{\i}n \v{C}ernoch}
\affiliation{Joint Laboratory of Optics of Palack\'{y} University and
     Institute of Physics of Academy of Sciences of the Czech Republic,
     17.~listopadu 50A, 779\,07 Olomouc, Czech Republic}
     
\author{Jan Soubusta}
\affiliation{Joint Laboratory of Optics of Palack\'{y} University and
     Institute of Physics of Academy of Sciences of the Czech Republic,
     17.~listopadu 50A, 779\,07 Olomouc, Czech Republic}
     
\author{Miloslav Du\v{s}ek}
\affiliation{Department of Optics, Palack\'y University,
     17.~listopadu 50, 772\,00 Olomouc, Czech~Republic}

\date{\today}

\begin{abstract}

The optical implementation of the recently proposed
unambiguous identification of coherent states is presented.
Our system works as a programmable discriminator between
two, in general non-orthogonal weak coherent states. The principle of
operation lies in the interference of three light beams --
two program states and one unknown coherent state which
can be equal to whichever of the two program states. The
experiment is based on fiber optics. Its results confirm theoretical predictions and 
the experimental setup can be straightforwardly extended for higher numbers of program
states.

\end{abstract}


\maketitle


\section{Introduction}

Only orthogonal states of any quantum system can be
discriminated perfectly and with a hundred per-cent
efficiency. However, it is possible to discriminate even
non-orthogonal states but either with errors and/or with
certain number of inconclusive results. In general, the
ability to discriminate quantum states is important for
quantum information transfer and processing. E.g., it can
serve as an efficient attack on quantum key distribution
\cite{DJL}. Error-prone discrimination was investigated
already in the seminal work of Helstrom \cite{helst}. Later
it was shown that the error-free or unambiguous
discrimination of two non-orthogonal states is also
possible though only in a probabilistic way
\cite{usd-I,usd-D,usd-P,j+s}. Unambiguous discrimination of
more than two non-orthogonal (but linearly independent)
states was also studied \cite{c+b}. The physical scheme for
optimal unambiguous discrimination of coherent states was
proposed by Banaszek \cite{ban}. Van Enk discussed an effectivity
of several methods for unambiguous discrimination of $N$ symmetric coherent 
states using linear optics and photodetectors \cite{enk}. Many other works dealt
with the discrimination of mixed states \cite{FJ,RLE,HB}.
Further, so-called programmable discriminators, where the set
of specimen states is determined by a quantum ``program'',
were proposed and experimentally tested
\cite{DuBu,exp-ol,BH}. This task can also be seen as
quantum state comparison \cite{J1}; the signal state is
compared with the set of specimen states. The
implementation of the coherent-state comparison device was
published by Erika Andersson  \emph{et al.} \cite{J2}.
Recently, Sedl\'{a}k \emph{et al.}
\cite{theory} proposed experimentally feasible
implementation of unambiguous identification of coherent
states with potential application to quantum database
search.

In this paper we present experimental realization of a
simple version of the unambiguous identification of coherent
states proposed in Ref.~\cite{theory}. The
uknown coherent state can equal to whichever of two different program states. 
The number of program states can be increased by extension of 
the basic experimental scheme.

\begin{figure}
  \begin{center}
    \smallskip
     \resizebox{\hsize}{!}{\includegraphics*{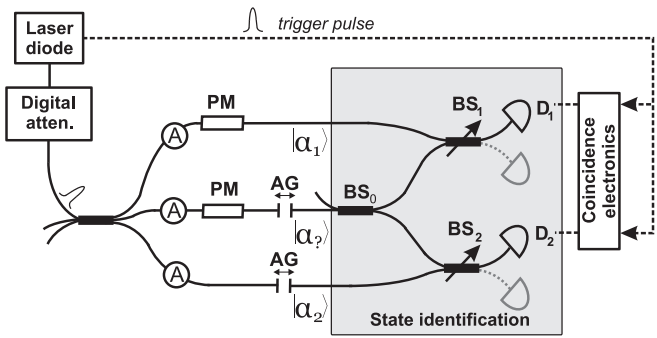}}
    \smallskip
  \end{center}
  \caption{The scheme of our experimental setup. A - attenuators, PM - phase modulators, AG - adjustable air gaps, BS - beam splitters, D - detectors.}
  \label{scheme}
\end{figure}

The scheme of our setup is in Fig.~\ref{scheme}. State
$|\alpha_?\rangle$ is the state to be discriminated. States
$|\alpha_1\rangle, |\alpha_2\rangle$ are the program
states. Our task is to find whether $|\alpha_?\rangle =
|\alpha_1\rangle$, or $|\alpha_?\rangle = |\alpha_2\rangle$.
As shown in Ref.~\cite{theory} if the intensity
transmittance of beam splitter BS$_1$,
\begin{equation}
  T_1=\frac{1}{1+T_0},
 \label{T1}
\end{equation}
and the transmittance of beam splitter BS$_2$,
\begin{equation}
  T_2=\frac{1-T_0}{2-T_0},
 \label{T2}
\end{equation}
where $T_0$ is the intensity transmittance of beam splitter
BS$_0$ (we suppose that the reflectances and
transmittances add to unity, $R_j + T_j = 1$), then one can
unambiguously identify the incoming state just by
photodetection at detectors D$_1$ and D$_2$. If D$_1$ clicks we can
conclude that $|\alpha_?\rangle = |\alpha_2\rangle$, if D$_2$
clicks it means that $|\alpha_?\rangle = |\alpha_1\rangle$.
If neither of the detectors clicks we cannot make any
conclusion about the state $|\alpha_?\rangle$ -- this
situation corresponds to an inconclusive result. In an
ideal situation both detectors may never click in coincidence. The
probability of correct identification of state
$|\alpha_1\rangle$ reads
\begin{equation}
  p_1 = 1 - \exp\left( -\eta_2\frac{1-T_0}{2-T_0} | \alpha_1-\alpha_2|^2 \right),
 \label{P1}
\end{equation}
and of correct identification of state $|\alpha_2\rangle$
\begin{equation}
  p_2 = 1 - \exp\left( -\eta_1\frac{T_0}{1+T_0} | \alpha_1-\alpha_2|^2 \right),
 \label{P2}
\end{equation}
where $\eta_j$ denote the detection efficiencies. In this paper we assume the equal prior probabilities of both coherent states $|\alpha_1\rangle$ and $|\alpha_2\rangle$. In such a case the optimal choice of the splitting ratios is shown to be $T_0=1/2, T_1=2/3, T_2=1/3$ (independently of the input states) \cite{theory}.

In a real setup it sometimes happens that detectors D$_1$ and D$_2$ click simultaneously. Practically these double clicks, as well as no detections, correspond to inconclusive results because we cannot distinguish whether the unknown state was equal to state $|\alpha_1\rangle$, or to state $|\alpha_2\rangle$. All other situations, when just one of detectors clicks, belong to conclusive results. They include both correct and erroneous identifications of the unknown state. Ideally, the erroneous identifications never occur and the probability of a conclusive result is equal to the probability of correct identification.


\begin{figure}
  \begin{center}
    \smallskip
     \resizebox{\hsize}{!}{\includegraphics*{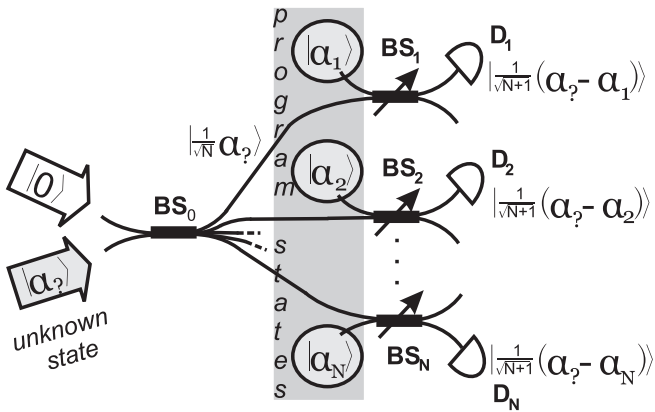}}
    \smallskip
  \end{center}
  \caption{Possible extension of the discrimination scheme for $N$ program states. BS$_0$ equally splits an unknown state to $N$ fibers. Splitting ratios of BS$_j$, $j=1,\ldots,N$:\quad $T_j=N/(N+1)$, $R_j=1/(N+1)$.}
  \label{scheme_n}
\end{figure}


\section{Description of experiment}

The experimental setup (see Fig.~\ref{scheme}) was built up on fiber optics. For preparation of coherent states we used  strongly attenuated pulses produced by a laser diode at wavelength 826~nm and with a length of pulses around 4~ns. Pulses were divided by a fiber coupler into three optical fibers, each corresponding to one of states  $|\alpha_1\rangle$,  $|\alpha_2\rangle$,  $|\alpha_?\rangle$. Amplitudes of coherent states were adjusted together by a digital attenuator located in front of pulse splitting and separately for individual modes by attenuators (A). Phases of states were controlled by electro-optical phase modulators (PM). 

The principle of state discrimination lies in the interference of light beams at beam splitters. The discrimination is optimal when beam splitter BS$_0$ is balanced, therefore a fiber coupler with fixed splitting ratio 50:50 was used. As beam splitters BS$_1$ and BS$_2$ we employed two variable-ratio couplers adjusted to desired splitting ratios.  The whole setup including both the preparation of all coherent states and the identification of unknown state worked basically as two interconnected Mach-Zehnder (MZ) interferometers. To accomplish discriminating operation visibilities of both MZ interferometers had to be maximized. This was provided by aligning polarizations and setting the same optical paths of arms corresponding to $|\alpha_1\rangle$ and $|\alpha_?\rangle$ and paths corresponding to $|\alpha_2\rangle$ and $|\alpha_?\rangle$. Before the measurement the path balance was roughly done by the help of two adjustable air-gaps and precisely by phase modulators during the measurement itself. 

Changes of temperature and temperature gradients cause changes of refractive index of optical fibers and thereby the drift of phase in time. To reduce the phase-drift effect we utilized thermal isolation of the setup in a polystyren box. Additionally an active stabilization was performed to compensate a residual phase drift. Before each three-second measurement step phase deviations from the balanced state were monitored simultaneously in both MZ interferometers and if necessary a proper correction was done by means of phase modulators \cite{kloner}.

The signal was detected by four Perkin-Elmer single-photon counting avalanche photodiodes. Two of them, D$_1$ and D$_2$, served for both the discrimination and active stabilization while two others were used only for the stabilization. To minimize the influence of dark counts of detectors on a measurement we counted coincidences between signals from detectors D$_1$, D$_2$ and pulses that triggered the laser diode. For this purposes coincidence electronics including time-to-amplitude convertors and single-channel analyzers were utilized. A coincidence window was set to the value of 8 ns. The mean number of dark counts in the coincidence window was approximately $4\times10^{-7}$ whereas the mean number of signal counts in the coincidence window in our experiment ranged from 0.002 to 0.7.


\begin{figure}[!t]
  \begin{center}
    \smallskip
    \resizebox{\hsize}{!}{\includegraphics{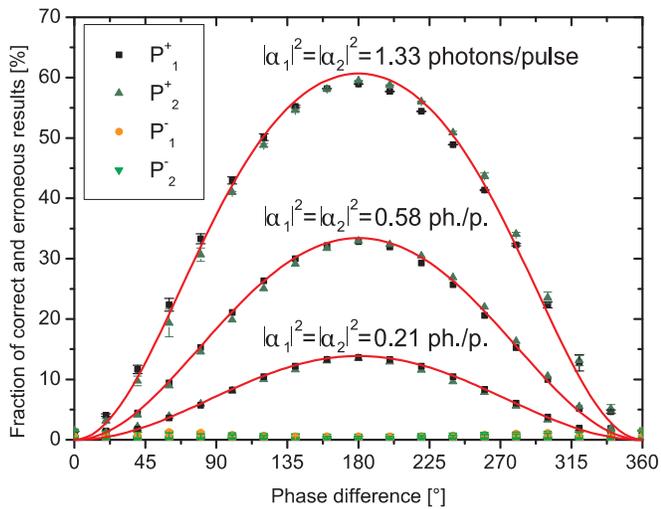}}
    \smallskip
  \end{center}
  \caption{Dependence of the fraction of correct and erroneous results on the phase difference between states $|\alpha_1\rangle$ and $|\alpha_2\rangle$ for three different intensities of states; $|\alpha_1|^2=|\alpha_2|^2$. Solid lines represent  theoretical predictions for the probability of a conclusive result.}
  \label{stejne}
\end{figure}



\begin{figure}[!t]
  \begin{center}
    \smallskip
    \resizebox{\hsize}{!}{\includegraphics{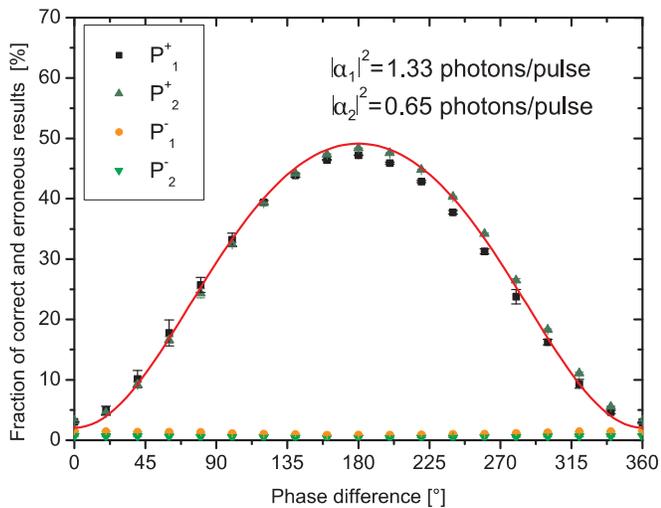}}
    \smallskip
  \end{center}
  \caption{Dependence of the fraction of correct and erroneous results on the phase difference between states $|\alpha_1\rangle$ and $|\alpha_2\rangle$; $|\alpha_1|^2\neq|\alpha_2|^2$. Solid line represents a theoretical prediction for the probability of a conclusive result.} 
  \label{ruzne}
\end{figure}


Quantum efficiences of detectors also play an important role. They are essential for the measurement of amplitudes of coherent states and they constrain the succes probability of state identification. Efficiencies were measured by means of a cw laser diode, a well-callibrated digital attenuator, and a power meter. First we determined the power of laser signal by the power meter. Then the laser beam was attenuated by the digital attenuator and the count rates measured by the detector were compared with a photon flux calculated from the power measured beforehand. By this measurement we obtained efficiences of detectors $\eta_1=\eta_2=53\pm1\%$.


\begin{figure}[!t]
  \begin{center}
    \smallskip
     \resizebox{\hsize}{!}{\includegraphics{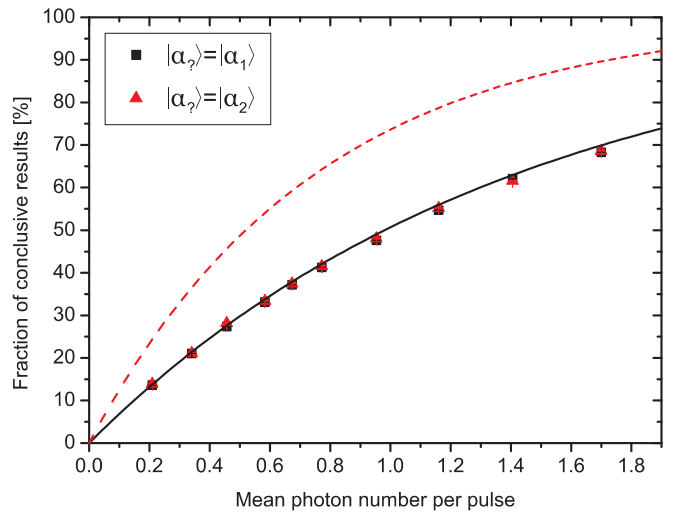}}
    \smallskip
  \end{center}
  \caption{Dependence of the probability of a conclusive result on the intensity of states ($|\alpha_1|^2=|\alpha_2|^2$; phase difference between states was 180$^{\circ}$). Solid line represents the theoretical prediction for our detectors with  $\eta=53\%$. Dashed line is the theoretical limit for ideal detectors (quantum efficiency $\eta=100\%$).}
  \label{Pmax}
\end{figure}



\begin{figure}[!t]
  \begin{center}
    \smallskip
    \resizebox{\hsize}{!}{\includegraphics{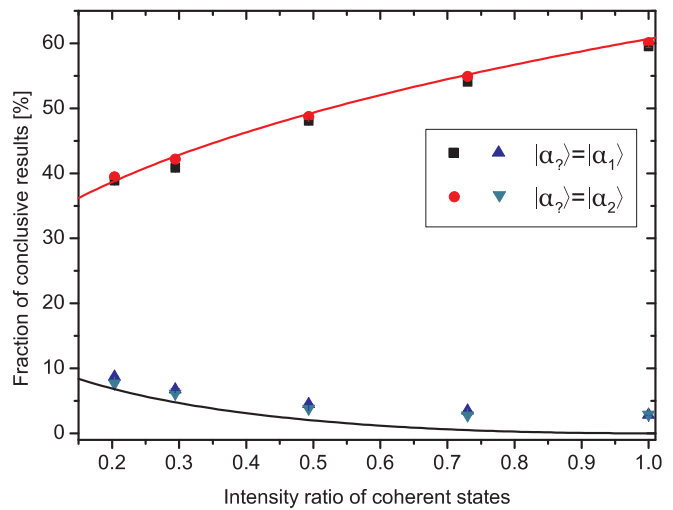}}
    \smallskip
  \end{center}
  \caption{Dependence of the probability of a conclusive result on the intensity ratio of states $|\alpha_2|^2/|\alpha_1|^2$ ($|\alpha_1|^2$=1.33 photons/pulse). The upper line represents the theoretical prediction for the phase difference 180$^{\circ}$ between states $|\alpha_1\rangle$ and $|\alpha_2\rangle$ and the lower line corresponds to the phase difference 0$^{\circ}$.}
  \label{Pmin}
\end{figure}


In our experiment we tested the state identification for various combinations of states $|\alpha_1\rangle$ and $|\alpha_2\rangle$. For each such measurement we first set desired intensities $|\alpha_1|^2$, $|\alpha_2|^2$ of program states. Then the intensity of state $|\alpha_?\rangle$ was adjusted to be equal either to the intensity of the first state or to the intensity of the second state. By applying proper voltages on phase modulators we were able to prepare coherent states with various phases. 

We measured conclusive count rates $C^{+}_j$, when the state was correctly discriminated, and $C^{-}_j$ related to erroneous detections; $j=1,2$. For example when $j=1$ then $|\alpha_?\rangle=|\alpha_{1}\rangle$. $C^{+}_1$ ($C^{-}_1$) were obtained by measuring coincidence rates between detector D$_2$ (D$_1$) and trigger pulses of laser diode minus the coincidence rates between detectors D$_1$ and D$_2$ (related to double clicks). $C^{+}_2$ and $C^{-}_2$ were measured in a similar way. The fractions of correct and erroneous results read
\begin{equation}
  P_j^+ = \frac{C_j^{+}}{C_{tot}},\qquad P_j^- = \frac{C_j^{-}}{C_{tot}}\qquad (j=1,2)
 \label{P_meas}
\end{equation}
respectively, where $C_{tot}$ is the total number of laser pulses per measurement period. The fraction of conclusive results is thus $P_j = P_j^{+}+P_j^{-}$ (j=1,~2).

\section{Results and conclusions}

Experimental results are shown in Figs.~\ref{stejne}-\ref{Pmin}. Each measured point was averaged from data collected during ten three-second measurements. Error bars correspond to statistical errors from these ten measuring steps. Fig.~\ref{stejne} and Fig.~\ref{ruzne} display the fraction of correct and erroneous results as a function of phase difference between coherent states $|\alpha_1\rangle$ and $|\alpha_2\rangle$. The theoretical curves of probabilities of a conclusive result (i.e.~probabilities of correct identification) were calculated by equations (\ref{P1}),~(\ref{P2}). In our case they are identical for the both program states due to the equality of the efficiencies of detectors D$_1$ and D$_2$. Measured data are presented as $P_j^{+}$ and $P_j^{-}$ according to equation (\ref{P_meas}). In the ideal case, when the visibility of interference is 100$\%$ and there are no dark counts, the probability of a conclusive result is equal to the probability of correct identification. In our setup the effect of dark counts was minimized to be negligible and visibilities were around 98$\%$. The imperfect interference affects the quality of discrimination mainly in situations when the overlap of coherent states $|\alpha_1\rangle$ and $|\alpha_2\rangle$ is relatively high.

The probability of a conclusive result for the phase difference 180$^{\circ}$ between states rapidly grows with increasing intensities of states $|\alpha_1\rangle$ and $|\alpha_2\rangle$ (see Fig.~\ref{Pmax}).

Fig.~\ref{Pmin} shows the probability of a conclusive result as a function of intensity ratio $|\alpha_2|^2/|\alpha_1|^2$ whereas the intensity of the first state was fixed to 1.33 photons per pulse. The upper line is related to situations when the overlap of states is for given intensities minimal (phase difference 180$^{\circ}$ between the states) and the lower line corresponds to cases when the overlap is maximal (phase difference 0$^{\circ}$ between the states).

Implemented version of coherent state identification can be straightforwardly extended for more than two specimen states (see Fig.~\ref{scheme_n}). There is no experimental limitation for the extension of this scheme and even the stabilization of more than two MZ interferometers can be performed simultaneously. However, the probability of correct identification decreases with increasing number of program states \cite{theory}.

In summary, we have experimentally demonstated the unambiguous programmable discriminator of coherent states when the uknown state can equal to one of two different specimen states. Measured values of fraction of correct results well agree with the theoretical predictions. Our experimental setup represents a technically feasible way to implement some interesting quantum information tasks, e.g. quantum database search \cite{theory} or an attack on quantum key distribution \cite{enk}.


\begin{acknowledgments}

M.D. acknowledges
fruitful discussions with Michal Sedl\'{a}k, M\'{a}rio Ziman,
Ond\v{r}ej P\v{r}ibyla, Jarom{\'\i}r Fiur\'{a}\v{s}ek, and Radim
Filip. This research was supported by the projects LC06007, 1M06002 and
MSM6198959213 of the Ministry of Education of the Czech Republic
and by the SECOQC project of the EC (IST-2002-506813).

\end{acknowledgments}


\end{document}